\documentstyle[eqsecnum,preprint,tighten,aps]{revtex}
\newfont{\largemi}{cmmi10}
\newfont{\smallmi}{cmmi6}

\begin{document}
\draft

\title{
Comment on ``Spectroscopy with Random and Displaced Random 
Ensemble"}

\author{ Y. M. Zhao$^{1}$ and  A. Arima$^{2}$}

\vspace{0.2in}
 \address{$^1$ Department of Physics,
 Saitama University, Saitama-shi, Saitama 338 Japan \\
$^2$ The House of Councilors, 2-1-1 Nagatacho, 
Chiyodaku, Tokyo 100-8962, Japan }

\maketitle

\date{\today}


C. W. Johnson and his collaborators recently found that 
the dominance of 0+ ground state (0 g.s.) dominance of even fermion 
systems can be obtained by using a two-body random ensemble (TBRE). 
This discovery  brought the physics of many-body 
systems interacting via random interactions into a sharp 
focus.  One of the most interesting and important issues 
is the origin of the 0 g.s.  dominance. 

The purpose of this Comment is to point out that 
the time reversal invariance, which was used in Ref. [2] 
to explain the 0 g.s. 
dominance, cannot be an explanation, and that the 
results by using a displaced TBRE are much more complicated than those 
 shown in  [2].  

The key point of [2] to explain the 0 g.s.  dominance 
is the behaviors of  centroid $E_{cI}$, defined as 
$d_I^{-1} tr(H)_I$, and the variance $\sigma_I^2$ 
defined as $d_I^{-1} \langle (H-E_{cI})^2 \rangle_I$. Here
$I$ is the total angular momentum. 
The $E_{cI}$ was noticed in Ref.[2] to be
  small. Therefore, the $\sigma_I^2$ was assumed to 
play  a crucial role.  This idea was actually
found to be not applicable to fermions in a single-$j$ shell 
in [3].  An argument, which is
essentially a combination of   behavior
of  $\sigma_I$ and a statistical behavior of two-body 
coefficients of fractional parentage, 
was  proposed  to explain both the properties  of  $E_{cI}$ and 
the 0 g.s. dominance [4] of even fermion systems in a single-$j$ shell.
A simple approach
to study this problem was   given recently in \cite{Zhao}. 

It is stressed here that the correlation between states 
is essential to explain the 0 g.s. dominance, or 
more generally, to explain a sizable probability
(denoted as $P(I)$)of a certain $I$
g.s. by using   a  TBRE.
Here correlation is an antonym  of independence, it 
refers to, e.g.,  for fermions in a single-$j$ shell,
the state with $I_{max}-2$ is  very likely the first
excited state when the $I=I_{max}$ state is the g.s..  
It is incorrect  to consider only 
the statistical behavior of energy levels. Taking $\sigma_I$ of 
5-fermion system in a single-$j$ shell as an example, $\sigma_I$
with $I=I_{min} = \frac{1}{2}$ is very  large. 
However, the $P(\frac{1}{2})$  is always
close to zero. One can also find many such examples
in even fermion systems.
The $\sigma_I$ of the $I=I_{max}$ state of fermions
in a single-$j$ shell  is always
0, but the $P(I_{max})$ is always sizable 
for fermions in a small $j$  shell. 

Next, the authors of Ref. [2] showed a few interesting examples 
by using  a  displaced TBRE. We emphasize here, however, that 
the results by using a displaced TBRE are actually very  
complicated. 
A negative displacement of the TBRE may favor 0 g.s., as 
showed in Ref. [2]. However,
this is not always correct. 
In fact, {\bf both} negative 
displacements and positive displacements may favor the 
0 g.s. probability (e.g., $P(0) \sim 100\%$ for 
  4 fermions in a two-$j$ ($j_1=13/2, j_2=9/2$) shell
by using a  TBRE$\pm 5$)
or quench down the
0 g.s. probability (e.g., $P(0) \sim 0\%$ for 
 4 fermions in a two-$j$ ($j_1=11/2, j_2=3/2$) shell
by using a  TBRE$\pm 5$), or may play a minor role
by a very slight change in $P(I)$'s
(e.g., 4 fermions in a two-$j$ ($j_1=9/2, j_2=5/2$) shell
by using a  TBRE$\pm 5$).

Our conclusion is that the time
reversal invariance, which was
further interpreted  in Ref. [2] by 
 the behaviors of    $E_{cI}$  and   $\sigma_I^2$, 
 can not be  the origin of the 0 g.s. dominance,
as was pointed out in \cite{Pittel}, and that 
the discussion by using  
a displaced TBRE in [2] is not true in  general.


\begin{thebibliography}{50}

\bibitem{Johnson1} C. W. Johnson, G. F. Bertsch, D. J. Dean, Phys. Rev. Lett.
{\bf 80}, 2749(1998).


\bibitem{Zuker} V. Velazquez, and A. P. Zuker, Phys. Rev. Lett. {\bf 88}, 
072502(2002).


\bibitem{Zhaox} Y.M. Zhao,  and A. Arima,  Phys. Rev. {\bf C64}, (R)041301(2001).


\bibitem{Arima} A. Arima, N. Yoshinaga, and Y.M. Zhao,  $International$ $Symposium$ $on$
$Nuclear$ $Structure$ $Physics$ 2001, World scientific (2001), P25, 
Edited by Rick Casten et al.; Eur. J. Phys. A, in press; 
N. Yoshinaga, A. Arima, and Y.M. Zhao, J. Phys. {\bf G}, to appear.


\bibitem{Zhao} Y.M. Zhao, A. Arima, and N. Yoshinaga, nucl-th/0112075; to be published.

\bibitem{Pittel} R. Bijker, A. Frank, and S. Pittel, Phys. Rev. {\bf C60},
021302(1999). 


\end{thebibliography}
\end{document}